%% file: paper9.tex
\documentclass[useAMS,usenatbib,twocolumn]{mn2e}
\usepackage{graphicx}
\usepackage{amssymb}
\usepackage{amsmath}
\usepackage{fleqn}
\usepackage{epstopdf}

\input{journal_defs}

\def\obs{{\boldsymbol{L}}}
\def\intd{{\mathrm{d}}}
\def\pos{{\boldsymbol{x}}}
\def\vel{{\boldsymbol{v}}}
\def\kpc{{\,\mathrm{kpc}}}
\def\params{{\mathcal{P}}}
\def\errmatrix{{\boldsymbol{C}}}
\def\Gauss{{\mathcal{G}}}
\def\kms{{\mathrm{kms^{-1}}}}
\def\kmssq{{\mathrm{km^2s^{-2}}}}
\def\data{{\mathcal{D}}}

\def\upartial{\partial}
\def\samplesize{15000}

\title[Is the Dark Halo of the Milky Way Prolate?] 
      {Is the Dark Halo of the Milky Way Prolate?}
\author[Bowden et al.]{A.~Bowden$^1$\thanks{E-mail: adb61,nwe,aamw3@ast.cam.ac.uk}, N.~W. Evans$^1$, A.~A. Williams$^1$
  \medskip
  \\$^1$Institute of Astronomy, University of Cambridge, Madingley Road,
       Cambridge, CB3 0HA, UK}
\begin{document}

\date{Accepted  Received ; in original form }

\pagerange{\pageref{firstpage}--\pageref{lastpage}} \pubyear{2015}

\maketitle

\label{firstpage}

\begin{abstract}
We introduce {\it the flattening equation}, which relates the shape of
the dark halo to the angular velocity dispersions and the density of a
tracer population of stars. It assumes spherical alignment of the
velocity dispersion tensor, as seen in the data on stellar halo stars
in the Milky Way. The angular anisotropy and gradients in the angular
velocity dispersions drive the solutions towards prolateness, whilst
the gradient in the stellar density is a competing effect favouring
oblateness.  We provide an efficient numerical algorithm to integrate
the flattening equation. Using tests on mock data, we show that the
there is a strong degeneracy between circular speed and flattening,
which can be circumvented with informative priors. Therefore, we
advocate the use of the flattening equation to test for oblateness or
prolateness, though the precise value of $q$ can only be measured with
the addition of the radial Jeans equation. We apply the flattening
equation to a sample extracted from the {\it Sloan Digital Sky Survey}
of $\sim \samplesize$ halo stars with full phase space information and
errors. We find that between Galactocentric radii of 5 and 10 kpc, the
shape of the dark halo is prolate, whilst even mildly oblate models
are disfavoured. Strongly oblate models are ruled out. Specifically,
for a logarithmic halo model, if the asymptotic circular speed $v_0$
lies between $210$ and 250 kms$^{-1}$, then we find the axis ratio of
the equipotentials $q$ satisfies $1.5 \lesssim q \lesssim 2$.
\end{abstract}


\section{Introduction}\label{sec:introduction}

The flattening of the Milky Way's dark halo is a subject of much
interest, but also alas much confusion.  The flaring of the HI gas
layer, the modelling of tidal tails (particularly of the Sagittarius
stream) and the kinematics of stars have all been used to determine
the shape.  The literature contains claims that the dark halo is
oblate~\citep{Ol00,Ko10,Bo15}, spherical~\citep{Fe06, Sm09a},
prolate~\citep{He04,Ba11} and triaxial~\citep{La10}. Not all of these
claims can be correct, even though the flattening could change with
radius~\citep[e.g.,][]{Ve13}.

The shape of the dark halo controls the kinematics of stellar halo
stars via the Jeans equations. In recent years, there has been a
growing concensus that the density of the stellar halo has an oblate
double-power law or Sersic profile structure~\citep[see
  e.g.,][]{Wa09,De11,PD15} with a flattening or axis ratio of $q_\star
\approx 0.6$. Once the density law is reliably known, the motions of
the stellar halo stars enable the gravity field, and hence the shape,
of the dark halo to be ascertained.

Data sets with full phase-space information for stars in the smooth
inner halo are beginning to become available. The largest to date has
been presented by \citet{Bo10}. Halo stars are extracted from the
Sloan Digital Sky Survey (SDSS) by combined color and metallicity
cuts, whilst SDSS spectroscopy is used for radial velocities and
Palomar Observatory Sky Survey (POSS) astrometry for proper
motions. This gives a sample of $\sim \samplesize$ halo stars with all
6 phase space coordinates.

\begin{figure*}
\includegraphics[width=1.8\columnwidth]{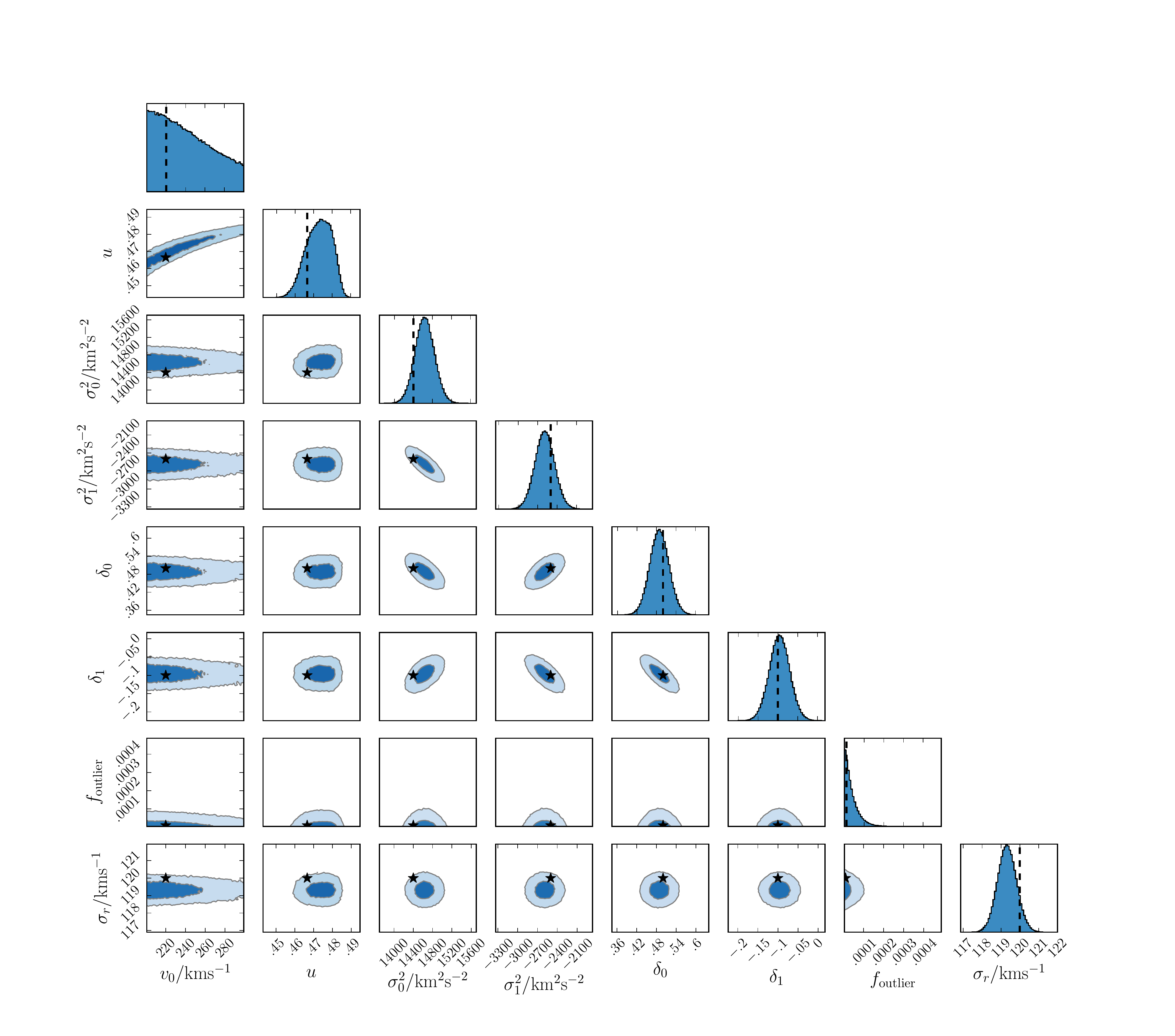}
\caption{One and two dimensional marginalisations of the posterior
  probability distribution when our method is applied to a mock
  catalogue of $\sim 24000$ stars without uncertainties. True values
  are marked by dashed lines (black stars) in the one (two)
  dimensional distributions. The nuisance parameters are all well
  constrained without bias. The degeneracy between $q$ and $v_0$ is
  apparent in the 2D marginalised distribution, though our prior on
  $v_0$ prevents the MCMC from sampling unrealistic circular speeds
  (and hence unrealistic values of $q$). The Gelman-Rubin statistic
  for this chain is $\simeq 1.002$.}
\label{fig:triangleplot}
\end{figure*}

The Jeans equations are often used in modelling external galaxies for
which only projected or line of sight data are usually
measurable~\citep[e.g.,][]{BDI,Ca08}. However, the situation
with the stellar halo of the Milky Way is rather different in that
line of sight velocities and proper motions are often available. All
the components of the stress or velocity dispersion tensor are in
principle measurable, as well as their spatial gradients.  This
motivates new approaches. For example, \citet{Lo14} have pioneered a
technique in which the data are used to derive acceleration maps
throughout the observed volume. By fitting the acceleration maps to
models, they find that the dark matter halo is very highly flattened
with an axis ratio of $q = 0.4 \pm 0.1$.  Here, we shall present
another technique which exploits the three dimensional nature of the
kinematic data now available for the Milky Way.

In principle, there are six components of the velocity dispersion
tensor, but only three Jeans equations.  However, if the alignment of
the velocity ellipsoid is known, this reduces to three equations for
the three unknown principal components. Here, we can exploit the fact
that the velocity ellipsoid of halo stars is closely aligned with the
spherical polar coordinate system with all the misalignment angles
small~\citep{Sm09a, Bo10, Ev15, Ki15}.  Spherically aligned solutions
of the Jeans equations have received little attention, despite early
work by \citet{Ba83} and \citet{Ba85}.  In Section 2, we derive the
angular Jeans equation -- or {\it the flattening equation} -- under
the assumption of spherical alignment. This relates the shape of the
dark halo to the angular velocity dispersions and the gradients in the
stellar density. Then, in Section 3, we devise a numerical algorithm
to solve the flattening equation and construct the likelihood. Section
4 shows tests of the procedure with mock data, together with
application to the dataset of \citet{Bo10} to deduce the shape of the
Milky Way's dark halo. We discuss our results, and compare to other
recent work, in Section 5.

\section{The Flattening Equation}
\label{sect:maths}

In an axisymmetric model, $\langle v_r v_\phi \rangle \equiv \langle
v_\theta v_\phi \rangle \equiv 0$ by the symmetries of the individual
orbits, which then leaves two non-trivial Jeans equations for the
tracer density $\nu (r,\theta)$ in terms of the potential
$\Phi(r,\theta)$:
\begin{eqnarray}
\null&& {\upartial \nu \langle v_r^2 \rangle \over \upartial r} 
 + {1 \over r} 
   {\upartial \nu \langle v_r v_\theta \rangle \over \upartial \theta}\nonumber\\
 &+& {\nu \over r } \Bigl( 2\langle v_r^2 \rangle 
             - \langle v_\theta^2 \rangle 
             - \langle v_\phi^2 \rangle 
             + \langle v_r v_\theta \rangle \cot\theta \Bigr)
   = \nu {\upartial\Phi \over \upartial r} \\
\null&&\null\nonumber \\
\null&& {\upartial \nu \langle v_r v_\theta \rangle \over \upartial r}
 + {1 \over r} 
   {\upartial \nu \langle v_\theta^2 \rangle \over \upartial \theta}\nonumber\\
 &+& {\nu \over r} \Bigl[ 3\langle v_r v_\theta \rangle 
             + (\langle v_\theta^2 \rangle 
                - \langle v_\phi^2 \rangle) \cot\theta \Bigr] 
  = {\nu \over r} {\upartial\Phi \over \upartial \theta}.
\end{eqnarray}
These two relations between the four stresses $\rho \langle v_r^2
\rangle$, $\rho \langle v_r v_\theta \rangle$, $\rho \langle
v_\theta^2 \rangle$, $\rho \langle v_\phi^2 \rangle$ must be satisfied
everywhere. We specialize immediately to spherical alignment, $\langle
v_rv_\theta \rangle =0$, as it holds good for the Milky Way's stellar
halo ~\citep[e.g.,][and references therein]{Ev15}.  This leaves two equations for
three unknowns $\langle v_r^2\rangle, \langle v_\theta^2 \rangle$ and
$\langle v_\phi^2 \rangle$, if the density and potential are given.
The first is the familiar radial Jeans equation:
\begin{gather}
 {\upartial \nu \langle v_r^2 \rangle \over \upartial r} 
  + {\nu \over r } \Bigl( 2\langle v_r^2 \rangle
  - \langle v_\theta^2 \rangle 
  - \langle v_\phi^2 \rangle  \Bigr)
  = \nu {\upartial\Phi \over \upartial r}.
\label{eq:radjeans}
\end{gather}
This equation is usually recast in a slightly different form, by
prescribing an anisotropy parameter $\beta(r)$ defined as $\beta(r) =
1 - (\langle v_\theta^2 \rangle +\langle v_\phi^2\rangle) / 2\langle
v_r^2 \rangle$.  Now the Jeans equation can be easily solved using the
integrating factor, as noted before many times
\citep[e.g.,][]{VdM94,Ma05,Ev06,Ag14}.

\begin{figure}
\includegraphics[width=0.9\columnwidth]{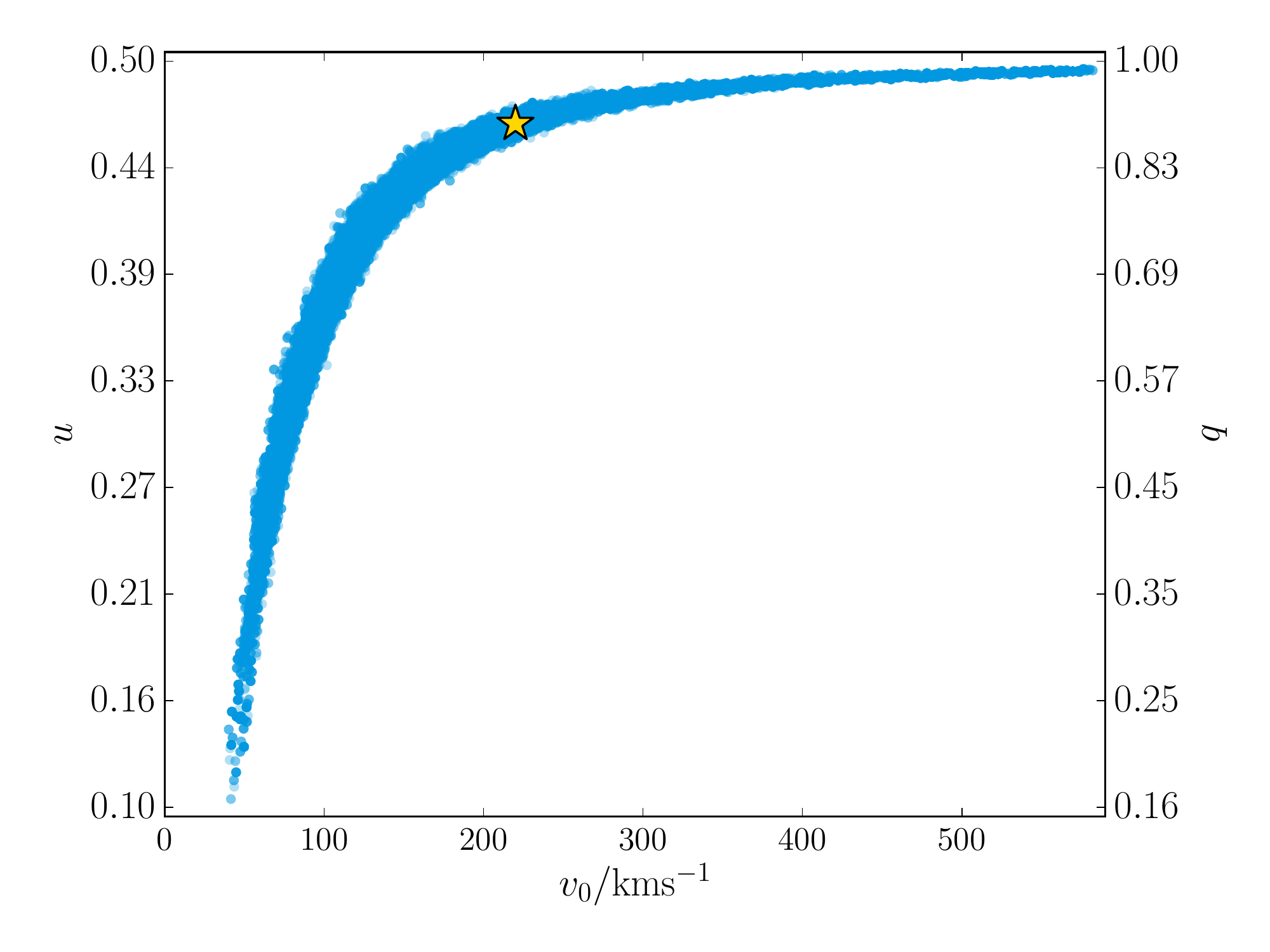}
\caption{Illustration of the degeneracy between the flattening $q$ (or
  $u$) and velocity normalisation $v_0$. We fix all model parameters
  to their input values except for the potential parameters, and then
  sample the remaining 2--dimensional space with {\sc emcee}. The
  input potential parameters are marked with a star. The posterior is
  almost constant along the centre of the degeneracy -- our method is
  unable to precisely constrain $q$ without prior knowledge of the
  normalisation of the potential.}
\label{fig:degeneracy}
\end{figure}

The second is the angular Jeans equation, or {\it the flattening equation}:
\begin{gather}
   {\upartial \nu \langle v_\theta^2 \rangle \over \upartial \theta}
 +  \nu  (\langle v_\theta^2 \rangle 
                - \langle v_\phi^2 \rangle) \cot\theta 
  = \nu {\upartial\Phi \over \upartial \theta}.
\label{eq:angjeans}
\end{gather}
The sign of the right-hand side of the equation is controlled by the
flattening of the gravity field, and is positive, zero or negative
according to whether the total matter distribution is oblate,
spherical or prolate.  For the Milky Way galaxy, the first term on the
left hand-side contains contributions from the angular gradient in
stellar density ${\partial \nu /\partial \theta}$. As the stellar halo
is oblate, this contribution is positive and drives the solution to
oblateness. However, the contribution from the angular gradient of
${\partial \langle v_\theta^2 \rangle /\partial \theta}$ is negative
and has the converse effect. Finally, the second term on the left-hand
side is small as $\langle v_\phi^2\rangle \gtrsim \langle v_\theta^2
\rangle$. It is usually negative (though not always, see e.g., Table 1
of Evans et al 2015) and so drives the solution to prolateness. The
result of the competing effects of these terms gives the flattening as
a function of radius.  Now notice something else that is simple, but
crucial. The flattening equation does not involve the radial velocity
dispersion at all. In spherical alignment, the flattening of the
potential is independent of both radial gradients and the radial
velocity dispersion. This decoupling considerably simplifies the
solutions of the axisymmetric Jeans equations.

First, though, we have to provide an analogue of the anisotropy
parameter $\beta(r)$ that prescribes how the angular velocity
dispersions are related. We must have $\langle v_\theta^2 \rangle =
\langle v_\phi^2 \rangle$ on the pole $\theta=0$ on symmetry grounds.
It has long been known that, away from the pole, $\langle
v_\phi^2\rangle \gtrsim \langle v_\theta^2 \rangle$ for the stellar
halo~\citep[see e.g.,][]{Ke07,Sm09b, Bo10}.  Without loss of
generality, the angular variation in the difference between $\langle
v_\theta^2 \rangle$ and $\langle v_\phi^2\rangle$ can be expanded in
an even Fourier series
\begin{gather}
1 - {\langle v_\phi^2 \rangle \over 
\langle v_\theta^2 \rangle} = -\sum _{m=1}^\infty\Delta_m
\sin^{2m} \theta,
\label{eq:fouriereven}
\end{gather}
where $\Delta_m$ may depend on $r$ but not on $\theta$.  This
naturally satisfies the constraint on the pole. Then, the flattening
equation is solvable with an integrating factor
\begin{gather}
j(\theta) = \prod_m \exp \left(-\frac{\Delta_m}{2m} \sin^{2m} \theta \right).
\end{gather}
If boundary conditions are given on the conical surface $\theta =
\theta_0$, then the full solution is
\begin{gather}
\nu \langle v_\theta^2 \rangle  = {1\over j(\theta)}
\Biggl[ j(\theta_0) \nu\langle v_\theta^2 \rangle \Biggl|_{r,\theta_0}
 + \int_{\theta_0}^\theta d\theta' j(\theta')\nu(r,\theta') {\partial
    \Phi\over \partial\theta'}\Biggr].
\label{eq:fullsol}
\end{gather}
This gives both $\langle v_\theta^2\rangle$ and $\langle
v_\phi^2\rangle$ everywhere as single quadratures.

Solutions of the Jeans equations for spherical alignment have been
given before~\citep{Ba83, Ba85, EHZ}.  The advantage of the algorithm
presented here is that it provides a practical parametrisation that
enables physically motivated solutions to the flattening equation to
be constructed very easily. As always with solutions of the Jeans
equations, there is no guarantee that there exists an underlying
positive-definite solution for the phase space distribution
function. In fact, \citet{Ev15} and \cite{An15} have shown that exact
spherical alignment of the second velocity moments implies that the
potential is separable or St\"ackel.  Nevertheless, even mild
misalignments allow a much greater range of possible equilibria, as
\citet{Ev15} demonstrated with made-to-measure N-body techniques. It
is reasonable to expect some, perhaps many, of our spherically aligned
Jeans solutions do therefore correspond to physical equilibria.

\begin{figure*}
\includegraphics[width=1.8\columnwidth]{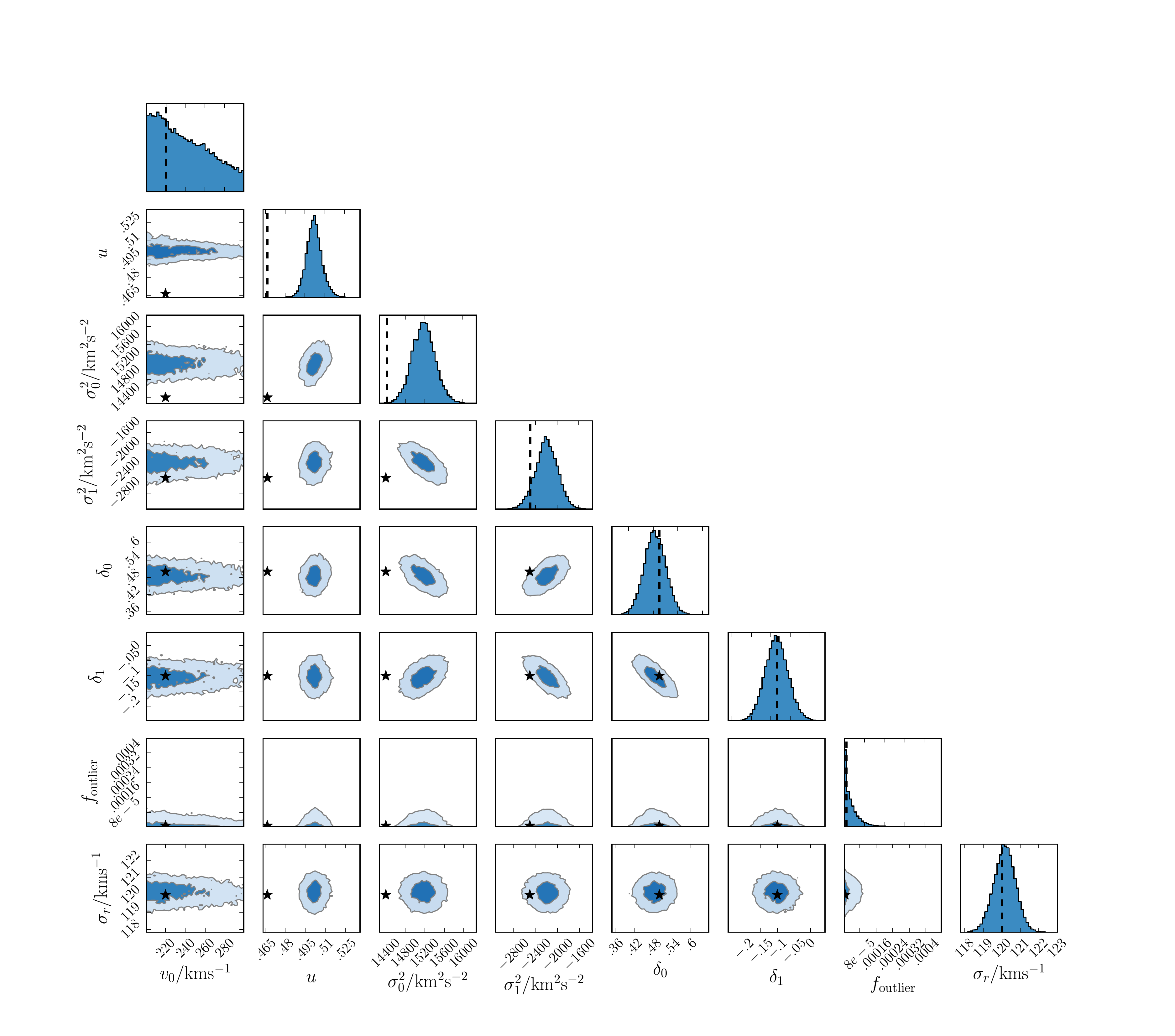}
\caption{As Fig.~\ref{fig:triangleplot}, but now using a mock
  catalogue of $\sim 24000$ stars incorporating uncertainties typical
  of the Bond et al. (2010) data.  The observational errors can
  scatter the value of the flattening away from the correct answer
  ($q=0.9$ or $u = 0.467$) to more spherical models. This is due to
  Poisson noise, which remains a problem despite the size of dataset.}
\label{fig:triangleplotwitherrors}
\end{figure*}

\begin{figure}
  \begin{center}
    \includegraphics[width=\columnwidth]{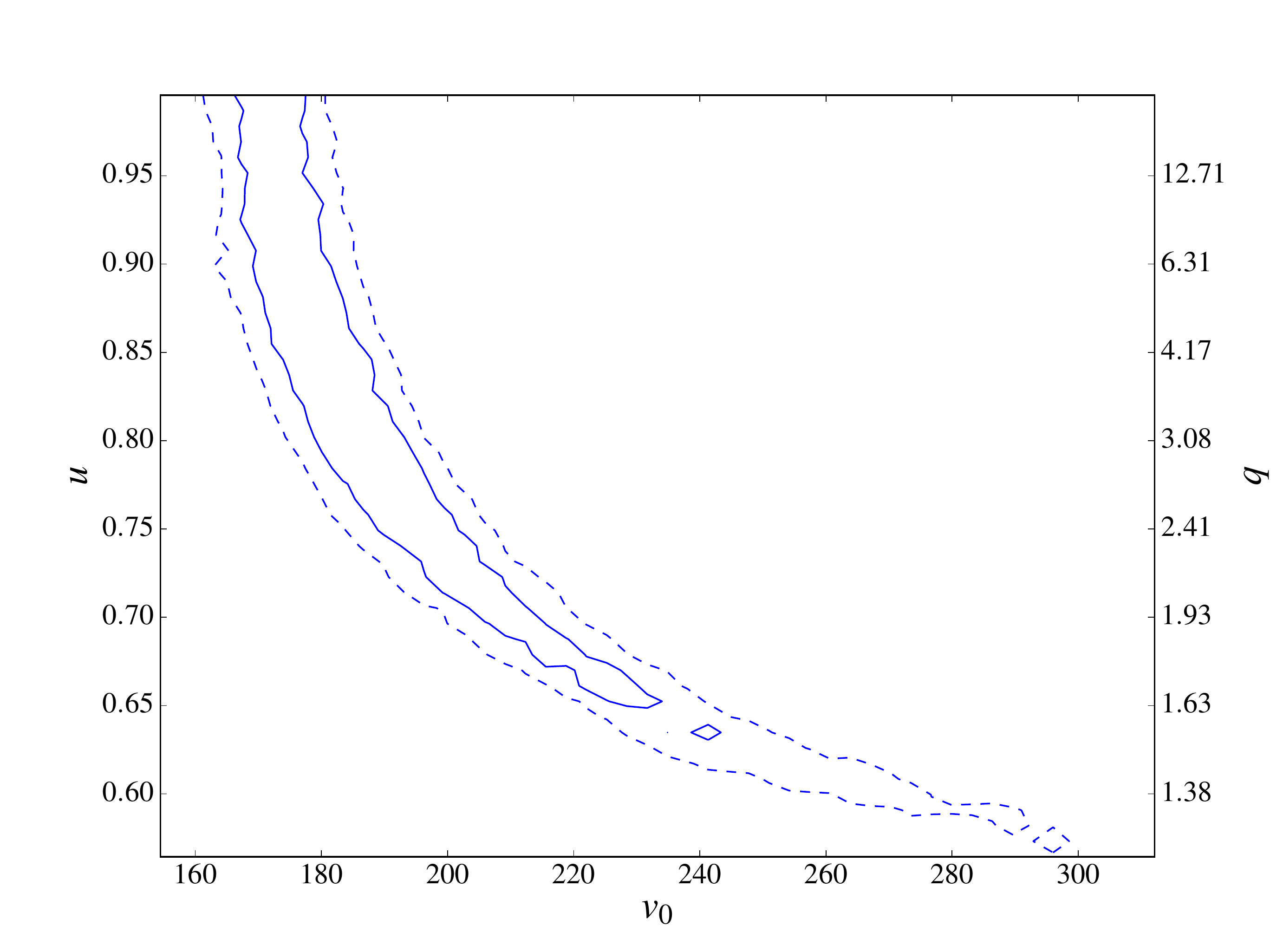}
  \end{center}
\caption{The two-dimensional posterior parameter distributions for
  velocity normalisation $v_0$ in kms$^{-1}$ and halo flattening $q$
  (or $u$) of 140,000 samples for the converged Monte Carlo chain. We
  observe the expected degeneracy here, however for the range of
  plausible halo normalizations the shape remains distinctly prolate.
  The Gelman-Rubin statistic for this chain is $\simeq 1.004$.}
\label{fig:sbsnodisk}
\end{figure}
\begin{figure}
  \begin{center}
    \includegraphics[width=\columnwidth]{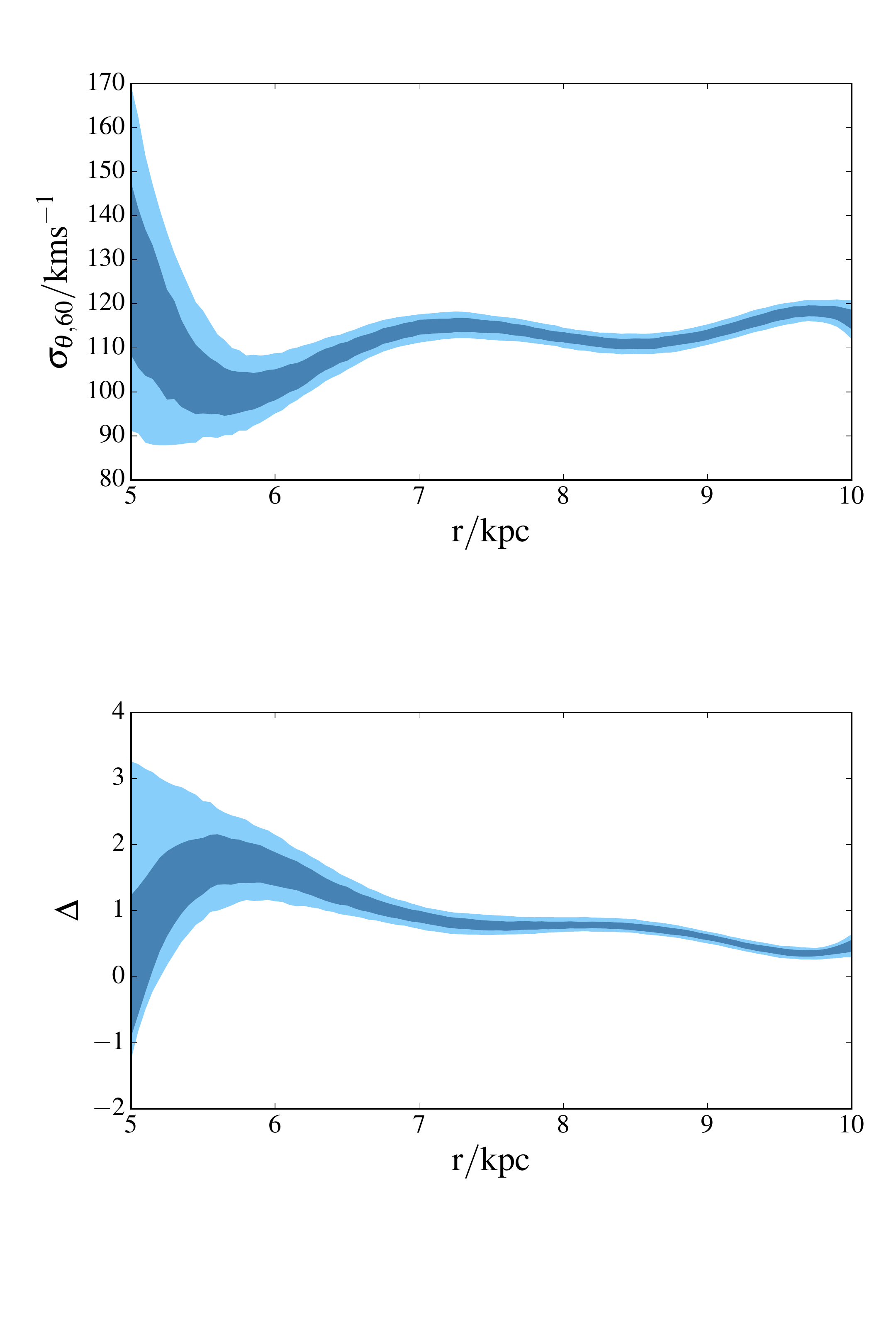}
  \end{center}
\caption{Plot showing an aggregation of posterior parameter
  distributions for 25,000 samples for the converged Monte Carlo
  chain. The dark blue shows 1$\sigma$ contours containing 68.27\% of
  the samples; the light blue shows 2$\sigma$, containing 95.45\%.The
  left panel shows the boundary condition for our integration, the
  value of $\sigma_\theta^2$ on the conical surface $\theta_0 =
  60^\circ$. The right panel shows the angular anisotropy parameter
  $\Delta$ parameter, which relates the two components of the angular
  velocity dispersion.}
\label{fig:bcs}
\end{figure}

\section{Numerical Implementation}

\subsection{Data}

We now describe the data set which we wish to model in a little more
detail.  \citet{Lo14} extract halo stars from the SDSS data by cutting
in colour $0.2 < g-r < 0.6$ and in metallicity [Fe/H] $< -1.1$.
There is some contamination from the disk populations, which is
estimated at $\sim 6$ per cent~\citep{Lo14}. To further reduce disk
contamination, we remove all stars with $|z| < 2$ kpc.

We choose to focus on a subsample extending over Galactocentric
spherical polar radii $5 \lesssim r \lesssim 10$ kpc. Requiring the
stars possess SDSS spectra for radial velocities and POSS astrometry
for proper motions yields a large sample of $14\,853$ halo stars. We
use the errors on the velocities computed by the SDSS pipeline, and
ignore any possible systematic effects.  The SDSS pipeline also
provides photometric errors, which can be used to compute random
errors in absolute magnitude (assuming that the covariances between
bands are negligible).  Polynomial relations for the absolute
magnitude of F and G stars as a function of $g$, $r$, $i$ and
metallicity [Fe/H] are given in \citet{Iv08}.

\subsection{Numerical Solution}

A significant defect of many solutions of the Jeans equations
presented in the literature is that the boundary conditions are often
applied at infinity. This requires assumptions as to the form of the
gravitational potential and density well beyond the limits of the
data, and so is unattractive. We eschew this practice and
integrate forwards and backwards from a cone embedded within the
data, using the data itself to set the boundary conditions.

We parameterize the angular velocity dispersion on the conical surface
$\theta= \theta_0$ as
\begin{gather}
\langle v_\theta^2 \rangle = \sum_m \sigma_m^2 \Big(\frac{r - r_{\rm
    d}}{\delta r}\Big)^m =
\sigma_{\rm d}^2(r),
\label{eq:boundarycondition}
\end{gather}
where $r_{\rm d} = 7.5$ kpc is a characteristic radius of the stars in
our sample and $\delta r = 2.5$ kpc is characteristic of the distance
range our data covers. This is of course a Taylor series about $r =
r_{\rm d}$.  We find that taking six terms of this series works well
in our application for which the data are restricted to spherical
polar radii in the range $5 \lesssim r \lesssim 10$ kpc.

The solution for the angular velocity dispersion is obtained by
integrating the flattening equation forwards or backwards:
\begin{eqnarray}
\nu \langle v_\theta^2 \rangle &=& \exp\left(\frac{\Delta}{2}\sin^2\theta\right) \Biggl[
\exp\left(-\frac{\Delta}{2}\sin^2\theta_0\right) \nu(r,\theta_0) \sigma_d^2(r)\nonumber\\
&+& \int_{\theta_0}^\theta
  d\theta' \exp \left(-\frac{\Delta}{2} \sin^2 \theta' \right)
  \nu(r,\theta') {\partial \Phi\over \partial\theta'}
  \Biggr].
\label{eq:simplesol}
\end{eqnarray}
This is simpler than the full solution (\ref{eq:fullsol}), as the
quality of the present data set only warrants the use of the first
term ($m=1$) in the Fourier series~(\ref{eq:fouriereven}). We have
retained only this term and dropped the subscript on $\Delta_1$.
Then, the physical meaning of $\Delta$ is that it is the maximum value
of the angular anisotropy $\langle v_\theta^2 \rangle/\langle v_\phi^2
\rangle -1 $, which occurs on the equatorial plane $\theta = \pi/2$.
Just as for the boundary condition, we use a Taylor series to describe
the radial variation of the anisotropy parameter,
\begin{gather}
\Delta(r) = \sum_m \delta_m \Big(\frac{r - r_{\rm d}}{\delta r}\Big)^m,
\label{eq:deltaexpansion}
\end{gather}
and usually take the first six terms of the series in our applications
to data.

Solution of Eq. ~(\ref{eq:simplesol}) also requires parameterization
of the potential. We use a family of halo potentials of the
form
\begin{gather}
 \Phi = -v_0^2 \ln \Bigl(R^2 + {z^2\over q^2}\Bigr),
\label{eq:hmodel}
\end{gather}
where ($R,z$) are cylindrical polar coordinates.  Although widely
used~\citep[e.g.][]{Ev93,BT,Bo15}, it is unlikely that the Galactic
potential is as simple as this.  We have experimented with inclusion
of a baryonic disk, but we find that this make little difference to
our conclusions on the flattening.  This is less surprising on
recalling that we are primarily interested in determining whether the
halo is oblate ($q<1$) or prolate ($q>1$), and less concerned about the
specific value of $q$.

\subsection{Likelihood Construction}

We implement a full Bayesian analysis of the data.  The data form a
vector of observables $\obs =
\left(s,\ell,b,v_{||},\boldsymbol{\mu}\right)$. Here, $(s,\ell,b)$ are
heliocentric distance, Galactic longitude and latitude, while
($v_{||},\boldsymbol{\mu})$, are line of sight velocity and proper
motion.  We must transform the likelihood function from observable
space into spherical polars $\left(\pos,\vel\right) =
\left(r,\theta,\phi,v_r,v_\theta,v_\phi\right)$.  This is achieved via
a Jacobian factor
\begin{equation}
p(\obs | \params) = p(\pos, \vel | \params)\,\left| \dfrac{\partial
  \boldsymbol{\left(\pos,\vel\right)}}{\partial \boldsymbol{\obs}}
\right|,
\end{equation}
where
\begin{eqnarray}
\dfrac{\partial \boldsymbol{\left(\pos,\vel\right)}}{\partial
  \boldsymbol{\obs}} &=& \dfrac{s^4 \cos b}{r^2 \sin \theta}.
\end{eqnarray}
We now construct a model that is compatible with our solutions to the
Jeans equations. Using the product rule for probabilities, we have
\begin{equation}
p(\pos, \vel | \params) = p(\vel | \pos, \params)\,p(\pos | \params),
\end{equation}
where $p(\vel | \pos, \params)$ is the local velocity distribution and
$p(\pos | \params)$ is proportional to the density law of the tracer,
in our case the stellar halo.  We take $p(\pos | \params) = 2\pi r^2
\sin\theta \nu(r,\theta)$ where
\begin{equation}
\nu(r,\theta) \propto r^{-n_\star}\left( \sin ^2 \theta + \cos ^2 \theta /q_\star^2 \right)^{-n_\star/2}
\label{eq:stellardensity}
\end{equation}
with $q_\star = 0.64$ and $n_\star = 2.77$. This single power-law
model was found by \citet{Ju08} to be a good match to SDSS
main-sequence turn-off halo stars and is valid in the distance regime
probed by the \citet{Bo10} data. In our work, we take the density as
fixed in the sense that we do not seek to constrain its parameters,
but it would be simple to incorporate into future work, provided that
the survey selection function is known. We then assume that the local
velocity distribution of halo stars is given by the product of three
Gaussian distributions
\begin{eqnarray}
p_{\mathrm{halo}}(\vel | \pos, \params) &=& \Gauss\left(v_r\,|\,0\,,\,\sigma_r\right)\,\Gauss\left(v_\theta\,|\,0\,,\,
								\sigma_\theta(r,\theta)\,\right)
                                                                \nonumber\\
&\,&\Gauss\left(v_\phi\,|\,0\,,\,\sigma_\phi(r,\theta)\,\right)
\end{eqnarray}
where we have introduced the shorthand
\begin{equation}
\Gauss(v | \mu, \sigma) = \dfrac{1}{\sqrt{2\pi\sigma^2}}\exp\left(\dfrac{-(v - \mu)^2}{2\sigma^2}\right),
\end{equation}
and $\sigma_\theta(r,\theta)$ and $\sigma_\phi(r,\theta)$ are the
solutions for the velocity dispersions given the current set of
parameters $\params$ and the Jeans equations. Note that we are
assuming that the stellar halo has no mean rotation~\citep[see
  e.g.,][]{Ke07,Sm09b}.  This parameterization encodes our solutions
to the Jeans equations, ensures spherical alignment of the velocity
ellipsoid, and gives a plausible shape to the local velocity
distributions.

For now, we make the crude assumption that the radial velocity
dispersion is constant with position -- this will not bias our
inference, although it is a signal that we are neglecting further
information that is encoded in the (spherical) radial velocities of
the stars. It is evidently possible to also solve the radial Jeans
equation, and thus introduce $\sigma_r(r,\theta)$, which would provide
further information about the gravitational potential. We opt not to
do this currently, since it involves the evaluation of a
two--dimensional quadrature and is hence much more time--consuming.
For the time being, we explore how much can be learned from
consideration of only the tangential velocities. To account for
outliers in the data, we add a further component to the velocity
distribution given by
\begin{equation}
p_\mathrm{outlier}(\vel) = \prod_{i=r,\theta,\phi} \Gauss(v_i | 0, 1000\kms)
\end{equation}
so that the full local velocity distribution is 
\begin{equation}
p(\vel | \pos, \params) = (1-f)p_\mathrm{halo} + fp_\mathrm{outlier}, 
\end{equation}
where $f$ is the outlier fraction, which we leave as a free parameter.

Given the above discussion, the likelihood function for an individual
datum with no observational errors is
\begin{equation}
p(\obs | \params) = p(\vel | \pos, \params)\, 2\pi s^4 \nu(r,\theta) |\cos b|.
\label{eq:likelihood}
\end{equation}
To incorporate observational errors into our model, we write
\begin{equation}
p(\obs | \params) = \int \, \intd \obs'\, p(\obs | \obs', \errmatrix)\,p(\obs' | \params),
\label{eq:errint}
\end{equation}
where $p(\obs | \obs', \errmatrix)$ is the probability of observing
the coordinates $\obs$ given some true coordinates $\obs'$. We
characterize this distribution as a multivariate Gaussian with
covariance matrix $\errmatrix$. We further assume that $\errmatrix$ is
diagonal, so that there are no covariances between uncertainties on
each of the observable coordinates. Eqn~(\ref{eq:errint}) cannot be
evaluated analytically, and so we evaluate it using the principle of
Monte--Carlo integration.  As in Evans et al. (2016; see also McMillan
\& Binney 2013), we draw $N$ samples from $p(\obs | \obs',
\errmatrix)$, and approximate eqn~(\ref{eq:errint}) by
\begin{equation}
p(\obs | \params) \simeq \frac{1}{N} \sum_{i=1}^N p(\obs'_i | \params),
\label{eq:approxlikelihood}
\end{equation}
where $p(\obs'_i | \params)$ is the RHS of eqn~(\ref{eq:likelihood})
evaluated for the ith draw from the error ellipsoid of the star. It is
important to note that we use the {\it same} Monte--Carlo samples for
each star for each new set of parameters $\params$, so that the
likelihood is not dominated by random noise (McMillan \& Binney 2013).
This means that, in practice, some of the terms in our likelihood
function become constant as a function of the parameters, so that
eqn~(\ref{eq:likelihood}) can be written as
\begin{equation}
p(\obs | \params) \propto p(\vel | \pos, \params).
\end{equation}
Terms dependent only on the positions of the stars are not variable
between likelihood evaluations, since we use fixed Monte Carlo samples
and do not vary the density law of the stars. As usual, we then
presume that all of our stars are independent observations, and so the
full likelihood is the product of eqn~(\ref{eq:approxlikelihood})
for all of the stars in our sample.

\subsection{Prior and Posterior Distributions}

For the circular velocity of the gravitational potential $v_0$, we
experimented with a number of different priors.  For tests using mock
data, we show that a restricted range for $v_0$ is necessary in order
to perform inference on the flattening; consequently, we use a
Jeffreys prior within the limits $200 < v_0 /\kms <300$.

For the axis ratio $q$, we first make the transformation
\begin{equation}
u = \frac{2}{\pi}\arctan q,
\label{eq:defnu}
\end{equation}
which maps the infinite domain $[0,\infty]$ to the domain $[0,1]$. The
case $u=1/2$ maps to $q=1$, so that oblate models correspond to the
range $0<u<1/2$ and prolate models are defined by $1/2<u<1$. In the
case of complete ignorance of the axis ratio of the halo, an
appropriate prior on $u$ is therefore uniform on the domain
$0<u<1$. We then sample the parameter $u$, which has a finite range. 
For the nuisance parameters, namely the coefficients in
Taylor expansions of the angular anisotropy $\Delta(r)$ and the
boundary condition on $\sigma_\theta^2(r,\theta=\pi/3)$, we assign
priors that are uniform in regimes where the full functions satisfy
\begin{eqnarray}
-4 &<& \Delta(r) < 4, \nonumber \\
10^2 &<& \sigma_\theta^2(r,\theta=\pi/3) / \kmssq < 700^2,
\end{eqnarray}
for $5 < r/\kpc < 10$. The prior on the outlier fraction $f$ is
uniform on the interval $0 < f< 1$. Later, we shall experiment with
other priors, in particular on the velocity normalization of the
gravitational potential. Given our priors, the full posterior
probability for our model is then given by Bayes' theorem
\begin{eqnarray}
p(\params | \data) = \dfrac{p(\data | \params)p(\params)}{p(\data)},
\end{eqnarray}
where $p(\data | \params)$ is the product of
eqn~(\ref{eq:approxlikelihood}) for all of the stars in our sample and
$p(\params)$ are the prior distributions discussed in this section.

\section{Applications}

\subsection{Tests on Mock Data}

To analyze the constraining power of our model, we now present tests
on mock catalogues of stars drawn from our model.  Since we are
assuming perfect knowledge of the tracer density, the precise
positions of the stars in our mock catalogues are not important. All
that we require is that the same density law is used to solve the
Jeans equations as appears in eqn~(\ref{eq:likelihood}).  To draw a mock
sample, we use the following procedure:

\begin{enumerate}
\item Select parameters $\params$ for the model, including all nuisance parameters.

\item At the position (as implied by the reported distance) of each
  star in our SDSS sample, solve the Jeans equations given $\params$
  for $\sigma_\theta(r_i,\theta_i)$, $\sigma_\phi(r_i,\theta_i)$.

\item For each star, draw each velocity $v_j$ ($j = r\,,\, \theta\,,\,
  \phi$) from Gaussian distributions with $\mu = 0$ and $\sigma =
  \sigma_j(r_i,\theta_i)$. In the radial case, $\sigma_r$ is a single
  parameter and independent of position.

\item For each star we now possess $(\pos_i, \vel_i) =
  (r_i,\theta_i,\phi_i,v_{r,i},v_{\theta,i},v_{\phi,i})$. Convert
  these coordinates into $\obs = (s_i, l_i, b_i, v_{||,i},
  \boldsymbol{\mu}_i)$.

\item If scattering through some set of uncertainties
  $\boldsymbol{\delta} = (\delta_s, \delta_v, \delta_{\boldsymbol{\mu}}
  )$, then for each star draw a new distance, line--of--sight velocity
  and set of proper motions from Gaussian distributions, e.g. $s'_i
  \sim \Gauss(s_i,\delta_s)$.
\end{enumerate}

Using the above prior probability distributions and likelihood, the
results for a sample of comparable size to the SDSS data but without
errors are shown in Fig.~\ref{fig:triangleplot}. We use two terms in
each of the Taylor expansions of eqns~(\ref{eq:boundarycondition}) and
(\ref{eq:deltaexpansion}). We do not include any outliers in our
tests. The nuisance parameters are all well constrained. There is a
noticeable degeneracy, however, between the potential parameters.

The degeneracy between $v_0$ and $q$ is made explicit in
Fig.~\ref{fig:degeneracy}.  To produce the figure, we fixed all the
nuisance parameters to their correct values, and then explored the
remaining two--dimensional space with {\sc emcee}, using the same mock
data set as in Fig.~\ref{fig:triangleplot}.  The degeneracy is narrow
and very curved. Small circular speeds correspond to extremely oblate
axis ratios, and larger circular speeds correspond to nearly spherical
potentials. When the halo is prolate, the degeneracy looks much the
same, except reflected about $q=1$ and $q\rightarrow\infty$ when
$v_0\rightarrow 0$.  This degeneracy arises because the flattening
equation only deals with the angular derivative of the potential. As a
result, any solution is invariant under the transformation
$\Phi(r,\theta) \rightarrow \Phi(r,\theta) + F(r)$, where $F(r)$ is
arbitrary. One way to break this degeneracy would be to consider the
radial Jeans equation, which includes a term involving the radial
derivative of the potential. However, as we have mentioned previously,
this comes at a large computational cost. Fortunately, at plausible
values of the circular speed ($v_0 > 200\kms$), the degeneracy is
quite flat as a function of $v_0$. In other words, $v_0$ is not well
constrained, but $q$ is. This is our reasoning for using a prior that
limits $v_0$ to lie in the range $200 < v_0 /\kms <300$. Once this
prior is used, we recover the axis ratio of the potential reasonably
well. 

When we use error-free data and fix all of the parameters except $q$ to 
their correct values, then there remains a range of $\sim 0.05$ in the maximum
posterior value of $q$ across 10 Monte Carlo samples. So, our results
remain susceptible to Poisson noise despite the large sample
size. When the observational errors are included, the results are
shown in Fig.~\ref{fig:triangleplotwitherrors}. Again, a sample of
similar size to the SDSS data is used, together with the error
distributions quoted by \citet{Bo10}. Now, the posterior distribution
of the flattening peaks close to spherical ($q=1$ or $u=0.5$) rather
than at the oblate model ($q=0.9$ or $u = 0.467$) used to generate the
mock data. This effect is again a manifestation of Poisson noise, and
may cause models that are close to spherical to be misclassified in
terms of prolateness or oblateness.

Our results suggest that the flattening equation returns some useful,
but limited, information given the data -- we can say with reasonable
confidence whether the potential is oblate or prolate, but not the
degree to which this is the case, unless we use an informative prior
on the potential normalization. Errors may cause mildly oblate or
prolate models to be misclassified, but this applies at most to the
range given by $0.9 \lesssim q \lesssim 1.1$, as judged from our
experimentation.

\begin{figure*}
  \begin{center}
    \includegraphics[width=1.8\columnwidth]{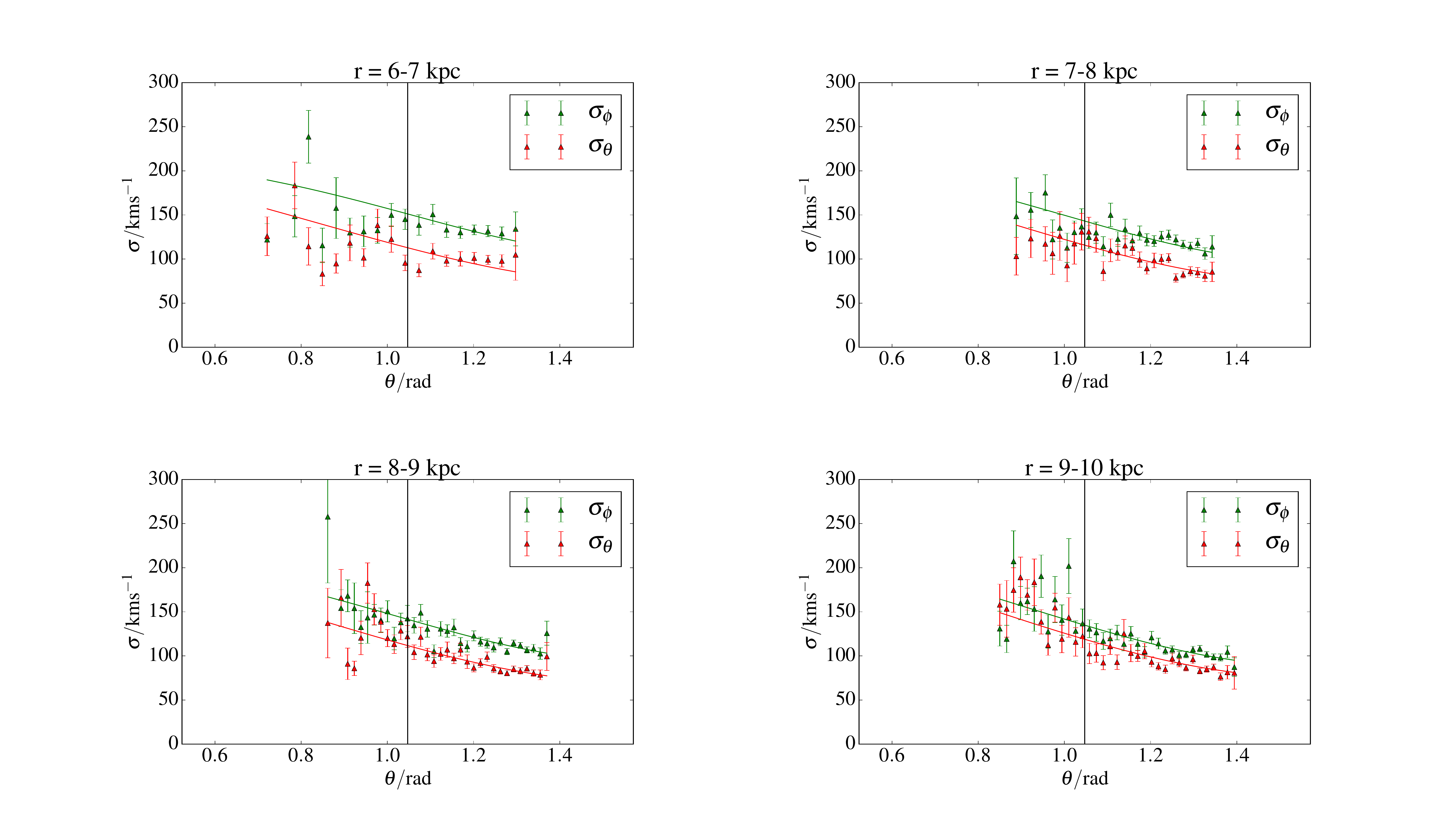}
  \end{center}
\caption{Sample fit for the binning method, showing the quality of the
  fit for the central three radial bins, namely 6-7 kpc (top left),
  7-8 kpc (top right), 8-9 kpc (bottom left) and 9-10 kpc (bottom
  right). Green shows the $\sigma_\phi$ component, red the
  $\sigma_\theta$. The vertical black line is marks the location
  $\theta_0 = 60^\circ$ where we set the boundary conditions for
  integration of the flattening equation.}
\label{fig:binsex}
\end{figure*}

\subsection{Results}

With the insights gained from mock catalogues, we now turn to the
data. We use the Bayesian method described above with the simple two
parameter halo consisting of a normalization $v_0$ and flattening $q$,
as given in eq~(\ref{eq:hmodel}). We use a uniform prior on $v_0$ and
a uniform prior on $u$ defined in eq~(\ref{eq:defnu}). There are
thirteen further nuisance parameters - six Taylor series terms giving
the boundary conditions for $\sigma_\theta$ on a conical surface, six
Taylor series terms describing the relationship between
$\sigma_\theta$ and $\sigma_\phi$, and one for $\sigma_r$. We used a
uniform prior on each of these parameters and assume no outliers in
this case.

Fig.~\ref{fig:sbsnodisk} shows the two-dimensional posterior parameter
distribution for $\sim$ 140,000 samples from a converged Monte Carlo
chain. We give the full one-dimensional posteriors for the velocity
normalization, flattening and all the nuisance parameters in Appendix
A.  We also performed the same fit with the inclusion of both a halo
and a razor-thin exponential disk. The disk had a central surface
density of $\Sigma_0 = 1.02 \times 10^9$ M$_{\odot}$kpc$^{-2}$ and a
scalelength $R_d$ = 2.4 kpc~\citep[c.f.][]{Bi12}. We verified that the
presence of a baryonic disk component has minimal influence on the
solution to the flattening equation in the region where we have data.

Although the data are not sufficient to constrain the halo flattening
precisely, we see that prolate haloes (i.e., $u> 1/2$) are strongly
favoured, with the posteriors ranging from near spherical to very
strongly prolate.  The flattening equation offers a robust
determination of the gross shape, as oblateness or prolateness
correspond to different signs of the left-hand side of
eqn~(\ref{eq:angjeans}). However, the actual constraint on the
numerical value of $q$ is weak. Stronger constraints could be obtained
by simultaneously solving both the flattening equation and the radial
Jeans equation (\ref{eq:radjeans}), but the computational cost is
punitive.  Stronger constraints could also be obtained with more
informative priors on $v_0$, similar to the ones used in the tests on
mock data.  Specifically, if the velocity normalisation $v_0$ lies
between $210$ and $250$ kms$^{-1}$ (as seems reasonable on other
grounds), then Fig.~\ref{fig:sbsnodisk} implies that the axis ratio of
the equipotentials $q$ satisfies $1.5 \lesssim q \lesssim 2$.  So, our
principal conclusion of prolateness does not change.  In
Fig.~\ref{fig:bcs}, we also show the constraints on both the angular
velocity dispersion on a conical surface $\theta_0 = 60^\circ$ above
the plane and the angular anisotropy parameter $\Delta$, which is
positive implying $\langle v_\phi^2\rangle > \langle v_\theta^2
\rangle$. These two parameters can be inferred directly from the data.

We can perform an alternative test of the quality of our fits using a
simple binning method. We divide our sample of stars into 2-D bins in
$r, \theta$. For each bin, we can evaluate $\sigma_i^2$ along with its
error and use the likelihood function
\begin{gather}
  \ln\mathcal{L} = -\frac{\chi^2}{2} = -\sum_{m} \frac{(\sigma_{\rm model,
      m}^2-\sigma_{i, m}^2)^2}{2\epsilon_{i,m}^2},
\end{gather}
where $\epsilon_{i,m}^2$ denotes the error in $\sigma_i^2$ in bin $m$.

In practice, the radial range $5 \le r \le 10$ is divided into 1 kpc
bins. The angular bins are chosen using the algorithm of \citet{Kn06},
with bins containing less than 10 stars removed. We find the results
are consistent with the probabilistic method. As the binning method
requires that we only draw once from the observational errors for each
star, we repeated the fits with a number of different random seeds.
Whilst there is a small amount of variation in the recovered
parameters, the results are broadly consistent at the 1$\sigma$
level. Fig.~\ref{fig:binsex} shows an example fit from the final chain
in the three central bins for the full probabilistic method; we see
that the model provides a resonable fit to the data in all four
bins. The reduced $\chi^2$ is 1.09 for this fit.

\section{Discussion and Conclusions}

The flattening equation robustly suggests that the shape of the Milky
Way's dark halo is prolate. Flattened oblate dark halos are strongly
disfavoured. These statements hold true in the inner halo over the
range of Galactocentric radii $5 \lesssim r \lesssim 10$ kpc.  If we
believe the halo is well represented by an axisymetric logarithmic
potential with velocity normalisation $v_0$ between $210$ and $250$
kms$^{-1}$, then the flattening equation implies that the axis ratio
of the equipotentials $q$ satisfies $1.5 \lesssim q \lesssim 2$.  This
result is consistent with the claim of \citet{Ba11}, who argued for
prolateness based on an analysis of the HI flaring gas layer from 8 to
24 kpc.

However, the result is at variance with the work of \citet{Lo14}, who
analyzed the same data of \citet{Bo10}.  \citet{Lo14} used the Jeans
equations in a novel way to generate two dimensional acceleration maps
from the data and then fit to galaxy models.  This has the advantage
that boundary conditions do not have to be prescribed, but it has the
disadvantage that spatial gradients of the density and the components
of the velocity dispersion tensor have to be evaluated from noisy
data.  They came to the the conclusion that the dark halo is highly
oblate with $b/a \approx 0.4 \pm 0.1$. Curiously, this is flatter than
the stellar halo itself, which has an axis ratio $b/a \approx 0.6$
\citep[e.g.,][]{Ju08,Wa09,De11,PD15}. As the dark matter is
dissipationless whereas the baryons are not, it is more natural to
expect that the dark halo is rounder than the stellar halo.

The flattening equation is a robust way to assess the oblateness or
prolateness of the dark halo. Although there is a strong degeneracy
between circular speed and flattening, we have shown that informative
priors on the circular speed enable the flattening to be recovered,
albeit with a mild bias when the data are scattered by the
observational errors. Therefore, we advocate the use of the flattening
equation to test for oblateness or prolateness, but not to measure the
precise value of $q$. If that is desired, the flattening equation must
be coupled with the radial Jeans equation in a computationally
efficient algorithm for large datasets. This work is under
investigation.

The only assumption in the method is spherical alignment of the
velocity ellipsoid, which has been confirmed by several independent
studies \citep{Sm09a,Bo10,Ev15,Ki15}. Other than three dimensional
velocity data, the main ingredient needed is the stellar halo density
law. This has also been measured by a number of workers~\cite[see
  e.g., Table 6 of ][]{PD15} with increasing consistency of results.
Although we used the law advocated by \cite{Ju08} here, we have
verified that different profiles \citep[e.g.,][]{De11} do not change
the principal conclusion of prolateness.

Other arguments have been advanced against oblate halos, of which the
most durable is the orbit of the Sagittarius stream. It would be
improper to suggest that this puzzling object is well understood, but
the conclusion that the orbit is inconsistent with oblate haloes has
persisted despite a wealth of new data and
modelling~\citep{Ib01,Ev14}. In fact, \citet{He04} already claimed
from analysis of the velocities of stars in the leading arm at $\sim
40$ kpc that the halo is prolate.  However, the stellar kinematical
analysis in this paper is confined mainly to the inner halo, whereas
the Sagittarius stream is a gigantic object with debris extending from
radii of 20 kpc to 100 kpc.

Of course, the coming of the first data sets from the {\it Gaia}
satellite will revolutionise this activity with abundant proper motion
data for halo stars brighter than $V\approx 20$. The flattening
equation will then provide a computationally efficient way of mapping
the halo shape as a function of Galactocentric radius using the space
velocities and errors of individual stars. This may finally resolve
the controversy of the shape of the Milky Way halo.

\section*{Acknowledgments}
We thank the anonymous referee for a thoughtful criticism of the first
version of this paper. ADB and AAW thank the Science and Technology
Facilities Council (STFC) of the United Kingdom for financial support.

\section*{Appendix A: Full Posteriors}

\begin{figure*}
  \begin{center}
    \includegraphics[width=2.2\columnwidth]{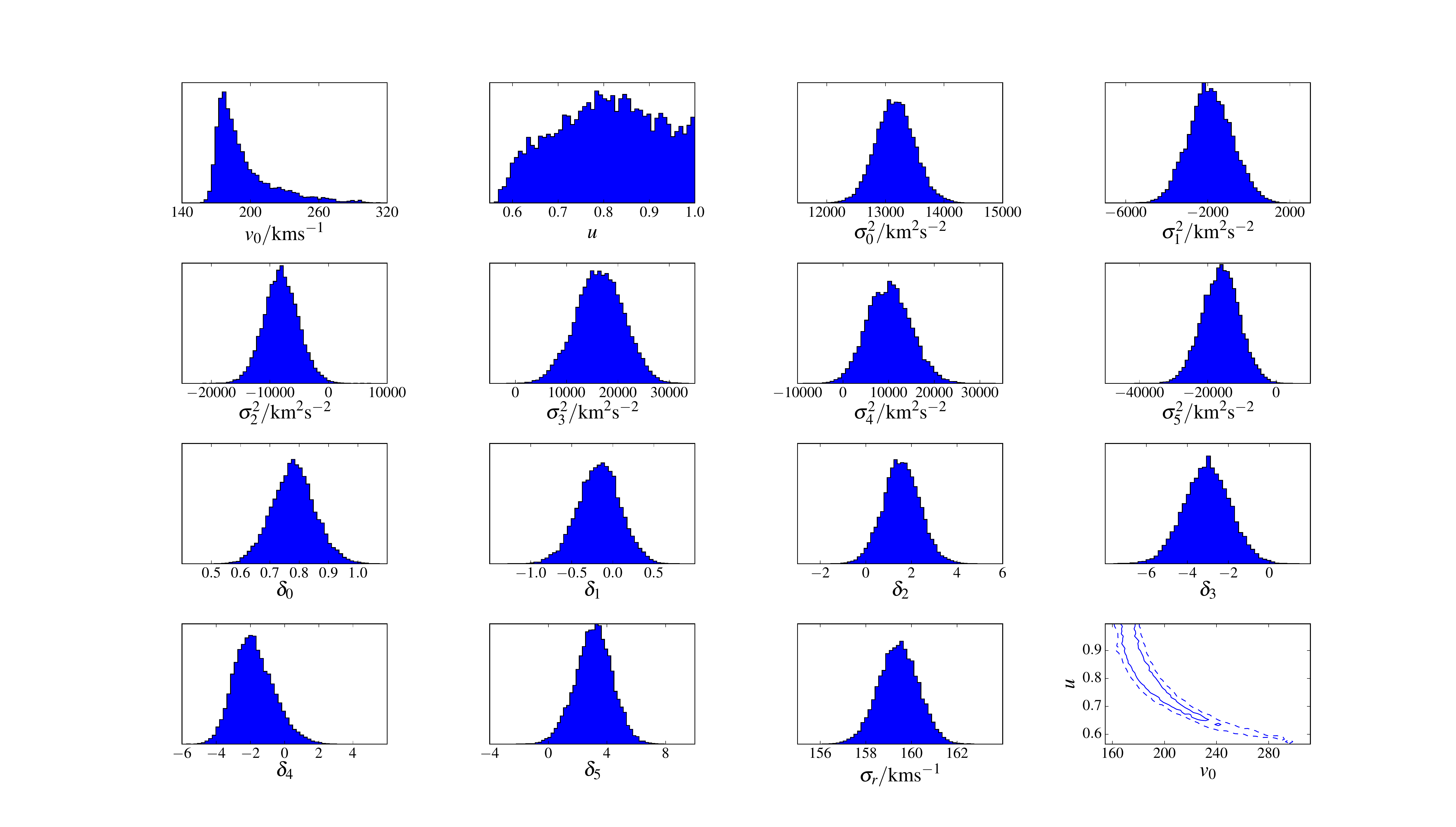}
  \end{center}
\caption{Plot showing the posterior parameter distributions of 140,000
  samples for the converged Monte Carlo chain. The fitting was
  perfomed using the method described in Section 3, with a two
  parameter halo model $(v_0,q)$. No baryonic disk component was
  included. The bottom right panel shows the two dimensional posterior
  for the halo parameters; we observe the expected degeneracy here,
  however even for realistic halo normalizations the shape remains
  distinctly prolate.}
\label{fig:sbsnodiskextra}
\end{figure*}

Here, for completeness, we give in Fig.\ref{fig:sbsnodiskextra} the
full posterior distributions on all the parameters for the final fit
to the Bond et al. (2010) sample.

\end{document}

%% file: journal_defs.tex
\long\def\symbolfootnote[#1]#2{\begingroup%
\def\thefootnote{\fnsymbol{footnote}}\footnote[#1]{#2}\endgroup} 
\def\aj{AJ}
\def\apj{ApJ}
\def\apjl{ApJ}
\def\aap{A\&A}
\def\mnras{MNRAS}
\def\pasp{PASP}